# Probing the Effects of Heat-Treatment Atmosphere on the Structural and Electrical Properties of NBT via $Eu^{3+}$ Photoluminescence

Zongxue Wang[1], Duanting Yan[1] and Hancheng Zhu[1,*]

[1]State Key Laboratory of Integrated Optoelectronics, Key Laboratory of UV Light-Emitting Materials and Technology of Ministry of Education, School of Physics, Northeast Normal University

Changchun, China

* Corresponding author: Hancheng Zhu (E-mail: zhuhc@nenu.edu.cn).

## Abstract

The effect of oxygen partial pressure during the pre-calcination step (high vacuum, air, nitrogen, and oxygen) on the crystal structure, microstructure, and electrical properties of $Na_{0.5}Bi_{0.465}Sr_{0.02}Eu_{0.005}TiO_{3-\delta}$ oxide-ion-conducting ceramics was systematically investigated. Dense ceramic samples were prepared by a conventional solid-state reaction route under different atmospheres. The results show that the oxygen partial pressure strongly affects Bi volatilization, grain growth, and oxygen-vacancy concentration. The largest average grain size (8.5 μm) was obtained for the nitrogen-treated sample, whereas the oxygen-treated sample exhibited the finest grains but the highest grain-boundary conductivity. X-ray diffraction and Raman spectroscopy indicate that low oxygen partial pressure enhances structural disorder, while high oxygen partial pressure stabilizes the lattice and promotes charge-transfer transitions. $Eu^{3+}$ photoluminescence further reveals the correlation between local structural evolution and oxygen-vacancy concentration. These findings clarify how oxygen partial pressure regulates grain-boundary behavior and ion-transport mechanisms in NBT-based oxide-ion conductors, providing guidance for optimizing the total conductivity of polycrystalline electrolytes.

## Introduction

Oxide-ion conductors are widely used in solid oxide fuel cells and high-temperature electrolysis cells for hydrogen production, and their future development requires both fabrication and operation at intermediate temperatures [1–5]. Sodium bismuth titanate (NBT) is a well-known ferroelectric and piezoelectric material that was first reported by Smolenskii and Agranovskaya in 1960 [6]. Previous studies have shown that the leakage behavior of NBT is closely related to the evaporation of a small amount of $Bi_2O_3$ during processing. Bi volatilization produces A-site deficiency and increases the oxygen-vacancy concentration in NBT [7,8]. It has also been reported that the electrical properties of NBT are highly sensitive to slight A-site non-stoichiometry. Bi-deficient sodium bismuth titanate is therefore considered a promising intermediate-temperature conductor. At 500 ℃, the bulk conductivity (σb) of $NB_{0.49}T$ doped with 2 mol% Sr or 1 mol% Mg is approximately $5.0 \times 10^{-3}$ S $cm^{-1}$, which is comparable to those of LSMG and GDC [9–11].

Most studies of NBT-based oxide-ion conductors have focused on enhancing bulk conductivity. In polycrystalline materials, however, the macroscopic total conductivity is governed jointly by the grain interior and the grain boundaries. When such materials are used in electrochemical devices, grain-boundary resistance becomes equally important. Grain-boundary resistance is affected by grain-boundary phase composition, grain size, impurity segregation, and space-charge layers, and it directly determines the macroscopic electrochemical performance of the material [12]. Cation doping and A-site non-stoichiometry can regulate the local $TiO_6$ octahedral framework, thereby affecting the distribution and migration of oxygen vacancies. $Sr^{2+}$, with a stable coordination environment and strong bonding ability in the perovskite lattice, is a suitable candidate for tailoring grain-boundary behavior in NBT-based oxide-ion conductors. Nevertheless, systematic studies on how oxygen partial pressure during pre-calcination affects the grain-boundary conductivity of Sr/Eu co-doped NBT-based ceramics remain limited. In this work, we investigate the effects of oxygen partial pressure on the structure, morphology, electrical properties, and photoluminescence behavior of $Na_{0.5}Bi_{0.465}Sr_{0.02}Eu_{0.005}TiO_{3-\delta}$ ceramics, with particular attention to ion transport in the grain-boundary region.

## Experimental

$Na_{0.5}Bi_{0.465}Sr_{0.02}Eu_{0.005}TiO_{3-\delta}$ ceramic samples were prepared by a conventional solid-state reaction method. $Na_2CO_3$, $Bi_2O_3$, $TiO_2$, $Eu_2O_3$, and $Sr(OH)_2$ were used as starting materials and weighed according to the stoichiometric composition. The powders were ground in anhydrous ethanol for 1 h and then pre-calcined at 800 ℃ for 2 h under high vacuum, air, nitrogen, or oxygen. After a second grinding step, the powders were pressed into pellets at 250 MPa and sintered at 1100 ℃ for 2 h in air. The samples are denoted as V-A, A-A, N-A, and O-A, respectively, where the first letter represents the pre-calcination atmosphere (vacuum, air, nitrogen, or oxygen) and the second letter indicates the final air-sintering atmosphere.

Phase identification was performed at room temperature using an X-ray diffractometer (Ultima IV, Rigaku, Japan) with Cu Kα radiation. The microstructure was examined by field-emission scanning electron microscopy (FE-SEM, Hitachi, Japan). Atomic-bond vibrations were characterized at room temperature using a Raman spectrometer (LabRAM HR Evolution, HORIBA, Japan) with 488 nm excitation. Silver paste was used as the electrode, and the silver-coated pellets were heat-treated at 800 ℃ for 0.5 h. AC impedance spectra were measured in air from 100 to 500 ℃ using an electrochemical workstation (CHI604E, CH Instruments, China) over a frequency range of 0.01 Hz to 1 MHz. A typical complex-impedance (Z*) plot of $Sr^{2+}$-doped $NB_{0.49}T$ contains three arcs, which correspond to the responses of the grain interior, grain boundaries, and electrodes from high to low frequency, respectively. The impedance data were fitted using an equivalent circuit consisting of three resistor-constant phase element (R-CPE) units connected in series. ZView software was used to extract the bulk and grain-boundary resistances and to calculate the corresponding conductivities.

## Results and discussion

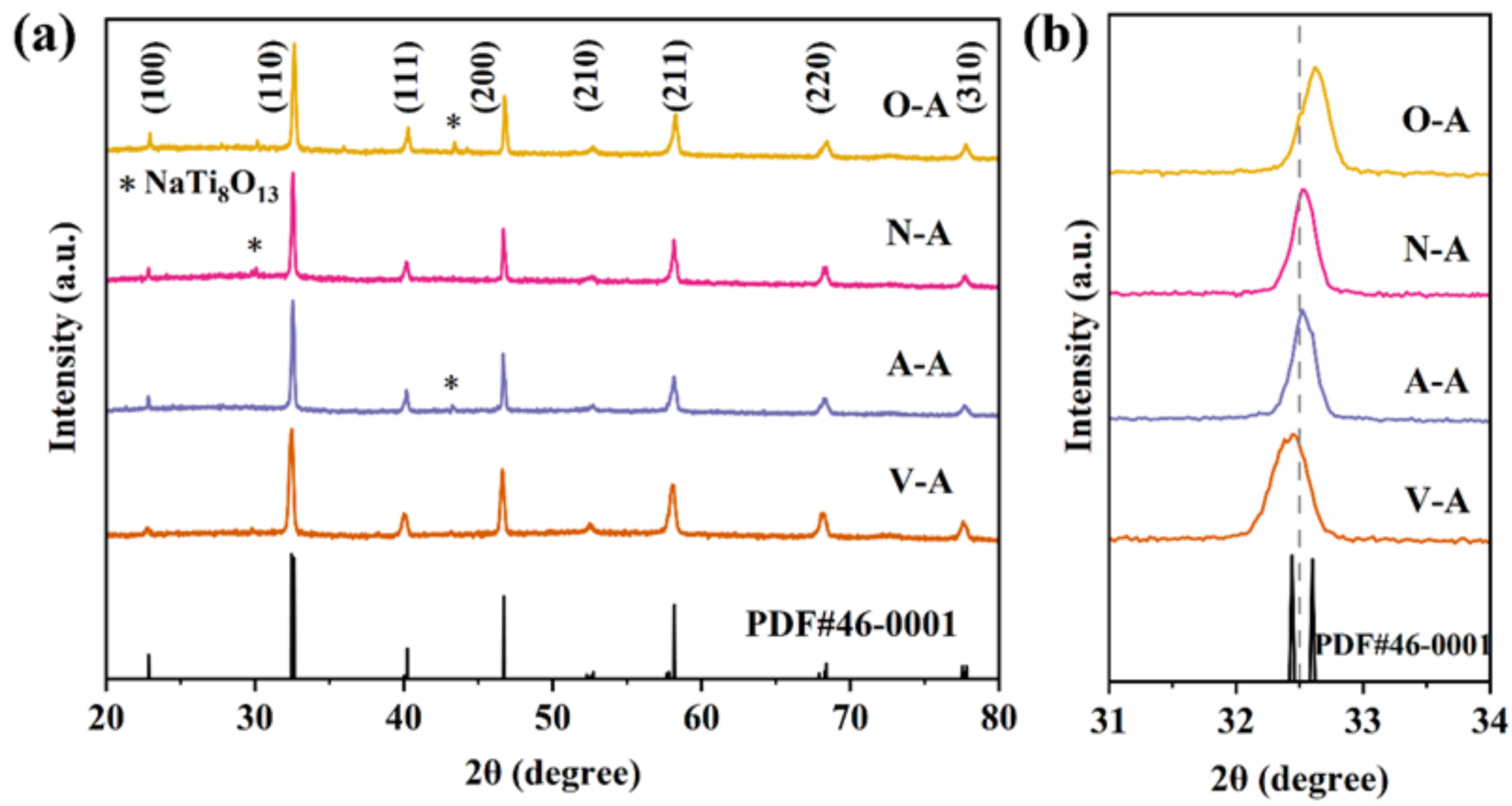

**Figure 1.** XRD patterns of $Na_{0.5}Bi_{0.465}Sr_{0.02}Eu_{0.005}TiO_{3-\delta}$ ceramics pre-calcined under different oxygen partial pressures: (a) full patterns and (b) enlarged view of the selected diffraction region.

Figure 1 shows that, except for the high-vacuum-pre-calcined sample, weak additional diffraction peaks are observed in the patterns. The marked reflections are assigned to $NaTi_8O_{13}$ (PDF#48-0523). The enlarged view on the right highlights the superlattice-related diffraction region. Although the vacuum-pre-calcined sample exhibits relatively higher phase purity, its crystallinity is poorer. This can be attributed to the enhanced volatilization of $Bi_2O_3$ under vacuum. Under such non-equilibrium conditions, Sr may not be homogeneously distributed, thereby disrupting the long-range ordering around the A-site cations.

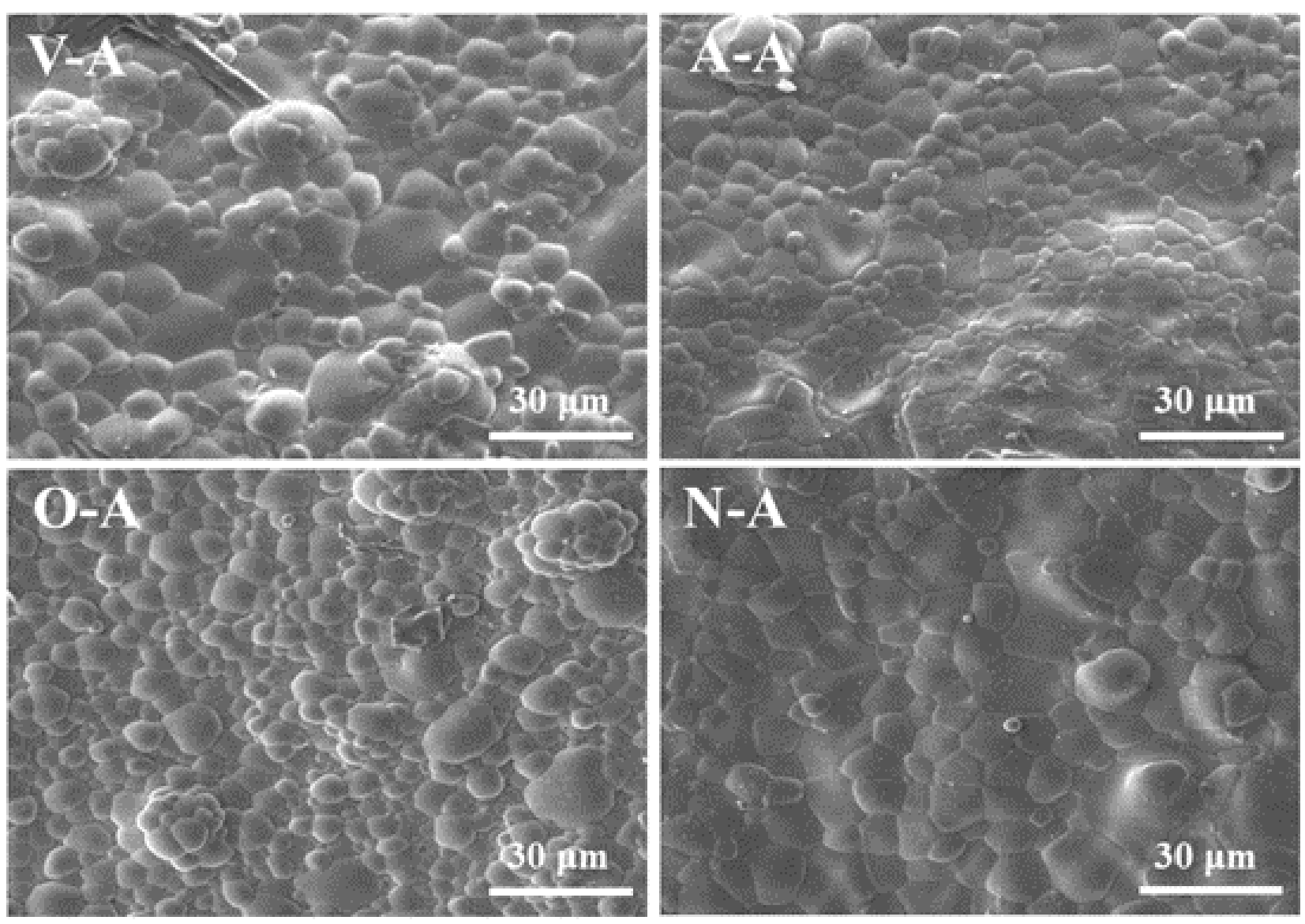

**Figure 2.** SEM micrographs of $Na_{0.5}Bi_{0.465}Sr_{0.02}Eu_{0.005}TiO_{3-\delta}$ ceramics pre-calcined under different oxygen partial pressures.

Figure 2 presents SEM images of the ceramic samples prepared under different oxygen partial pressures. All samples are dense and show uniform grain growth. The N-A sample exhibits the largest grain size, approximately 8.5 μm. A nitrogen atmosphere is generally considered inert or weakly reducing, and its oxygen partial pressure is much lower than that of air or oxygen. Under this condition, a higher concentration of oxygen vacancies is generated by lattice-oxygen loss, which facilitates ion migration and accelerates grain-boundary migration. At the same time, reduced oxygen adsorption at the grain boundaries weakens grain-boundary pinning and promotes the coalescence of small grains into larger grains. In contrast, under high oxygen partial pressure, extensive oxygen adsorption at the grain boundaries hinders their migration, leading to grain refinement in the O-A sample.

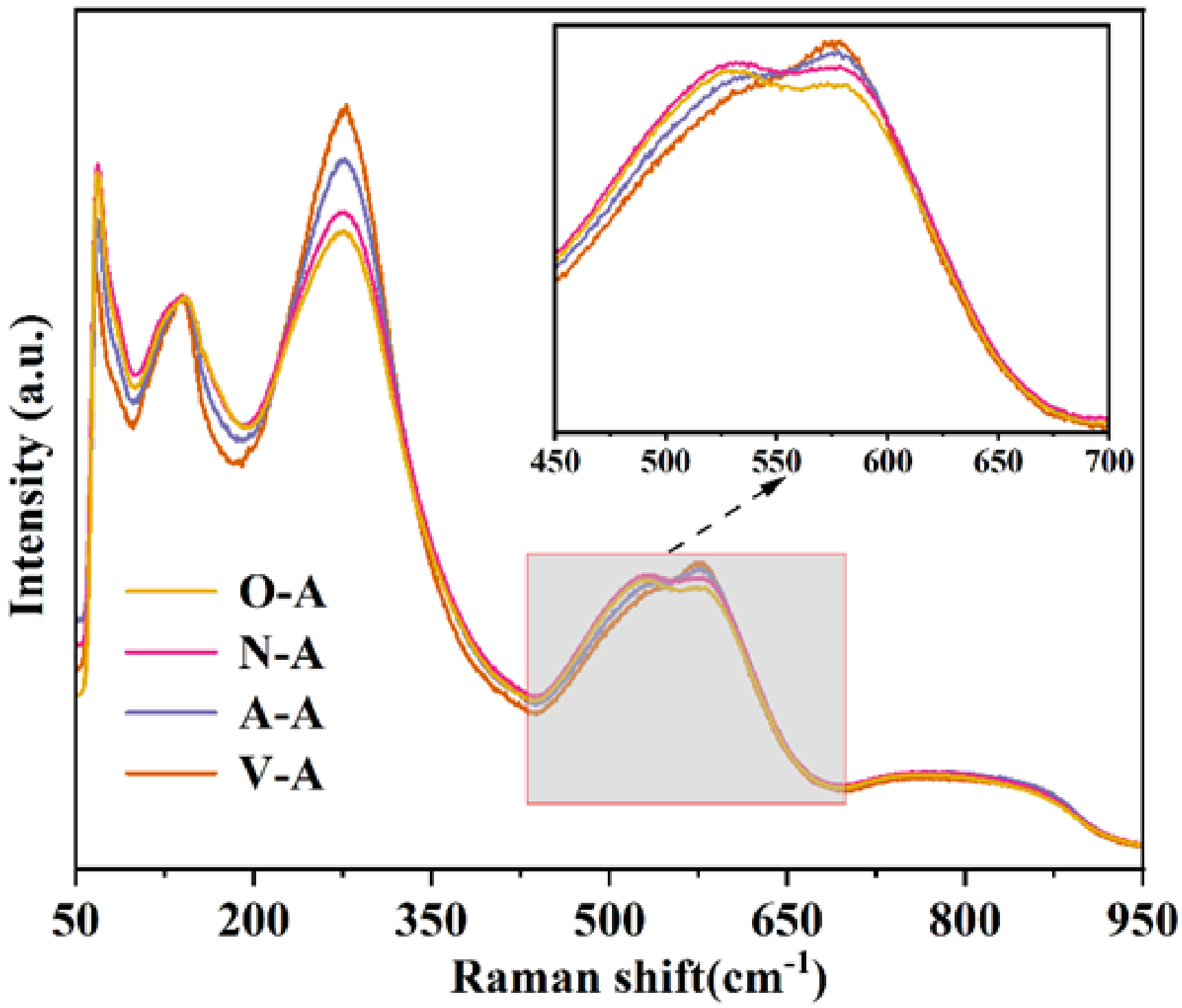

**Figure 3.** Raman spectra of $Na_{0.5}Bi_{0.465}Sr_{0.02}Eu_{0.005}TiO_{3-\delta}$ ceramics pre-calcined under different oxygen partial pressures.

The Raman spectra in Figure 3 show that the main Raman vibration bands of all samples are consistent with those of pure NBT, suggesting that no obvious impurity phase is detected by Raman spectroscopy. All spectra were normalized to the Na-O vibration band near 250 $cm^{-1}$. Under high oxygen partial pressure, oxygen-vacancy formation is suppressed and $Ti^{4+}$ is stabilized, preventing its reduction to $Ti^{3+}$. The more ordered local environment weakens the Ti-O bond response, which is also consistent with the sharpest diffraction peaks observed for the O-A sample in the XRD patterns. When the oxygen partial pressure decreases, oxygen vacancies are generated, inducing local distortion and lattice stress and destroying the uniformity of the vibrational modes; consequently, the environment around $Ti^{4+}$ in the N-A sample becomes more disordered. When the oxygen partial pressure is further reduced to a near-vacuum condition, $Bi_2O_3$ volatilization is intensified, the A-site stoichiometry deviates strongly from the nominal composition, and a large number of defects are introduced by lattice-oxygen loss. The local distortion is therefore the most severe and the structure becomes highly disordered. The pronounced change in the $TiO_6$ octahedral band shape of the V-A sample provides clear evidence for the disruption of long-range ordering, and the sub-band at lower wavenumber almost disappears.

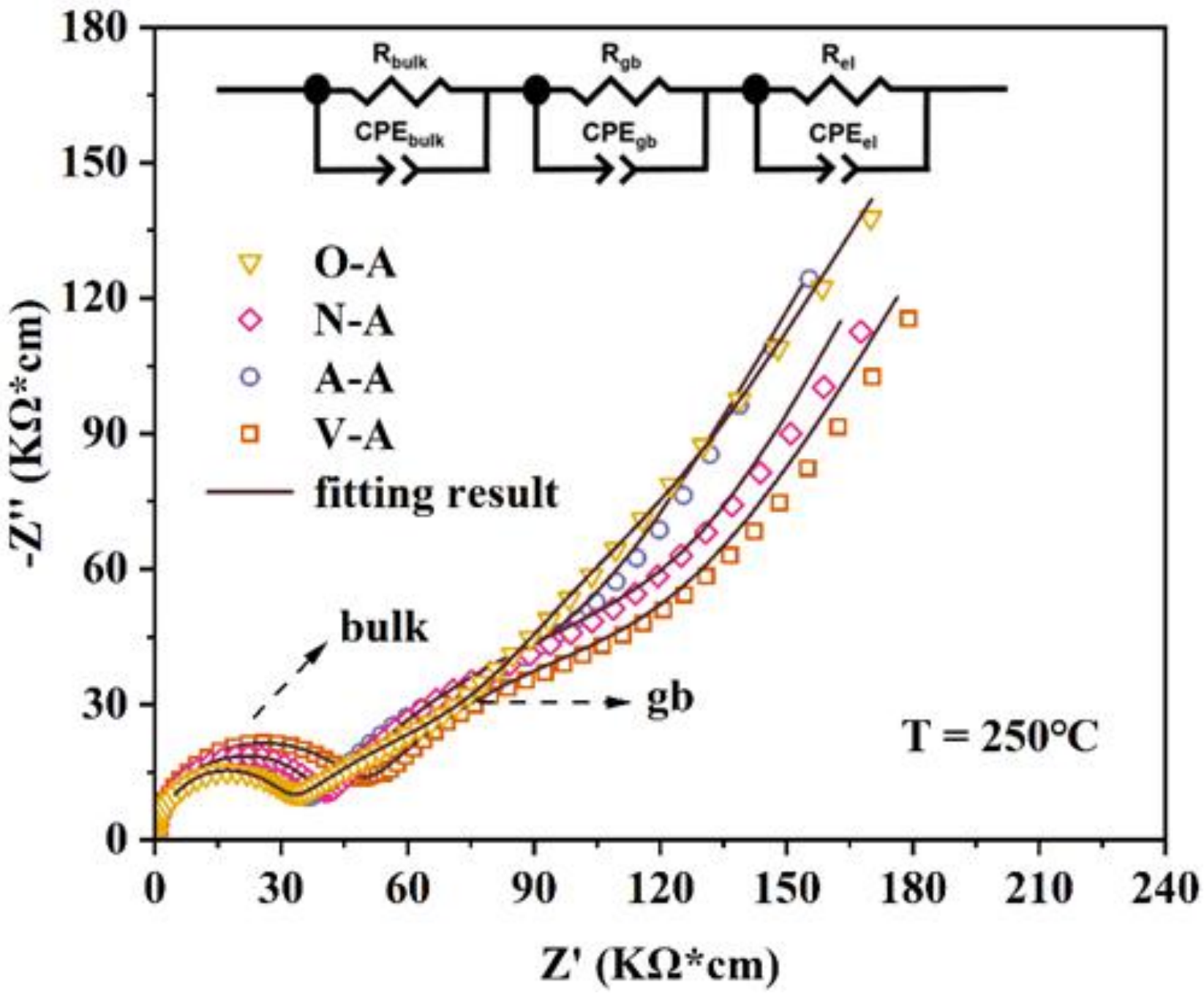

**Figure 4.** AC impedance spectra of $Na_{0.5}Bi_{0.465}Sr_{0.02}Eu_{0.005}TiO_{3-\delta}$ ceramics pre-calcined under different oxygen partial pressures.

Figure 4 shows representative AC impedance spectra of $Na_{0.5}Bi_{0.465}Sr_{0.02}Eu_{0.005}TiO_{3-\delta}$ ceramics measured at 250 ℃. The solid lines represent equivalent-circuit fitting results obtained using ZView, and they agree well with the experimental data. The high-, medium-, and low-frequency regions correspond to three arcs associated with the grain interior, grain boundaries, and electrode response, respectively. The bulk and grain-boundary resistances were obtained by fitting the impedance spectra, and the corresponding conductivities were calculated from the fitted resistance values. The bulk conductivity was calculated according to:

$$\sigma_{bulk} = L / (S \times R_{bulk})$$

where L is the pellet thickness, S is the electrode area coated with silver paste, and $R_{bulk}$ is the fitted bulk resistance obtained from ZView. The fitting results indicate that the O-A sample has the smallest bulk arc, whereas the V-A sample exhibits the largest bulk and grain-boundary resistances.

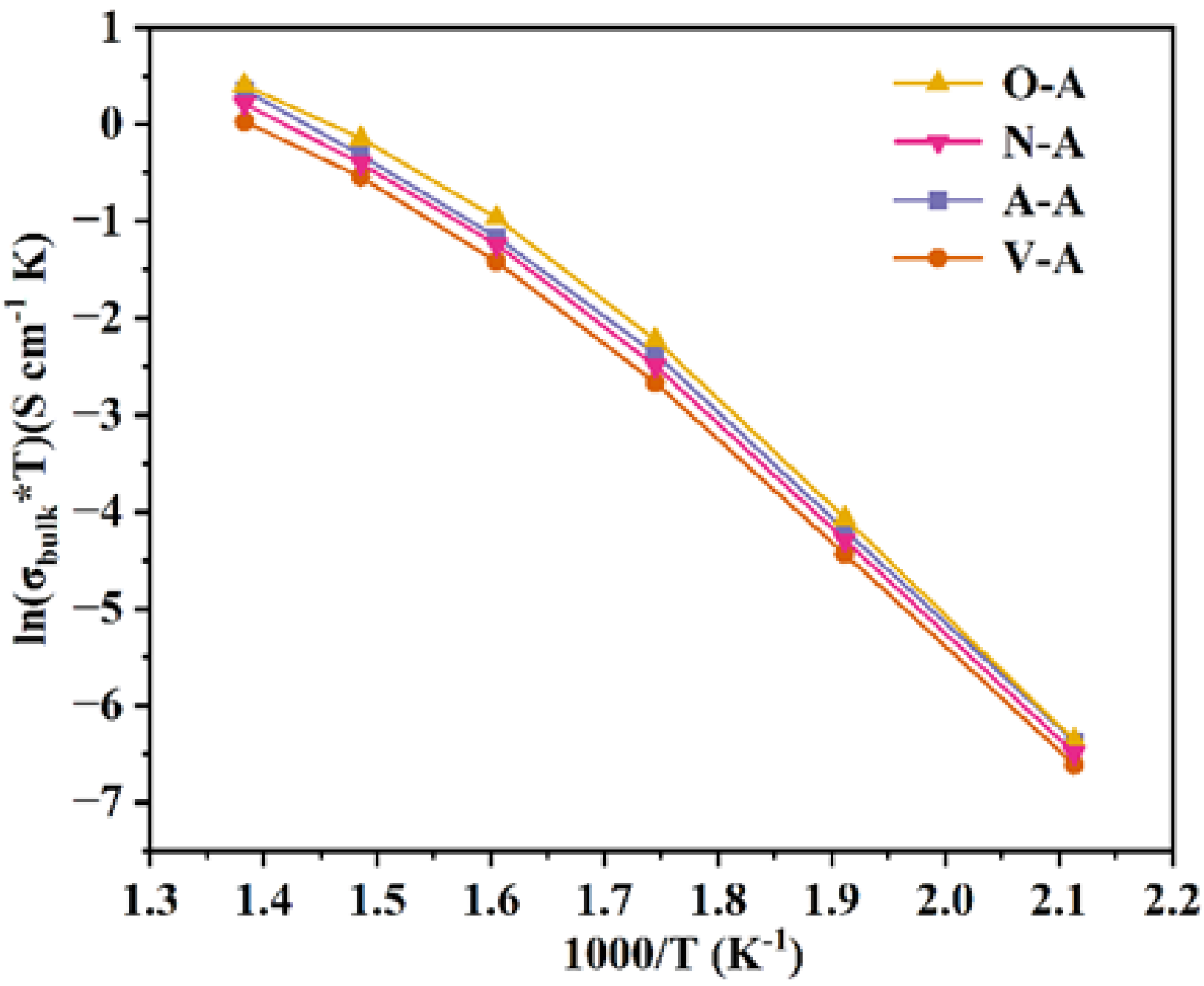

**Figure 5.** Arrhenius plots of bulk (grain) conductivity for $Na_{0.5}Bi_{0.465}Sr_{0.02}Eu_{0.005}TiO_{3-\delta}$ ceramics pre-calcined under different oxygen partial pressures.

Figure 5 shows the Arrhenius plots of the bulk conductivity of the ceramic samples prepared under different oxygen partial pressures. The O-A sample exhibits the highest bulk conductivity. As discussed above, the oxygen-vacancy concentration is expected to remain relatively low under high oxygen partial pressure; therefore, the enhanced bulk conductivity cannot be explained simply by an increased carrier concentration. Conversely, the V-A sample, which is expected to have the highest oxygen-vacancy concentration, shows the lowest bulk conductivity. The N-A sample, also prepared under low oxygen partial pressure, exhibits lower bulk conductivity than the A-A reference sample. These results indicate that differences in activation energy are the dominant factor responsible for the variation in conductivity.

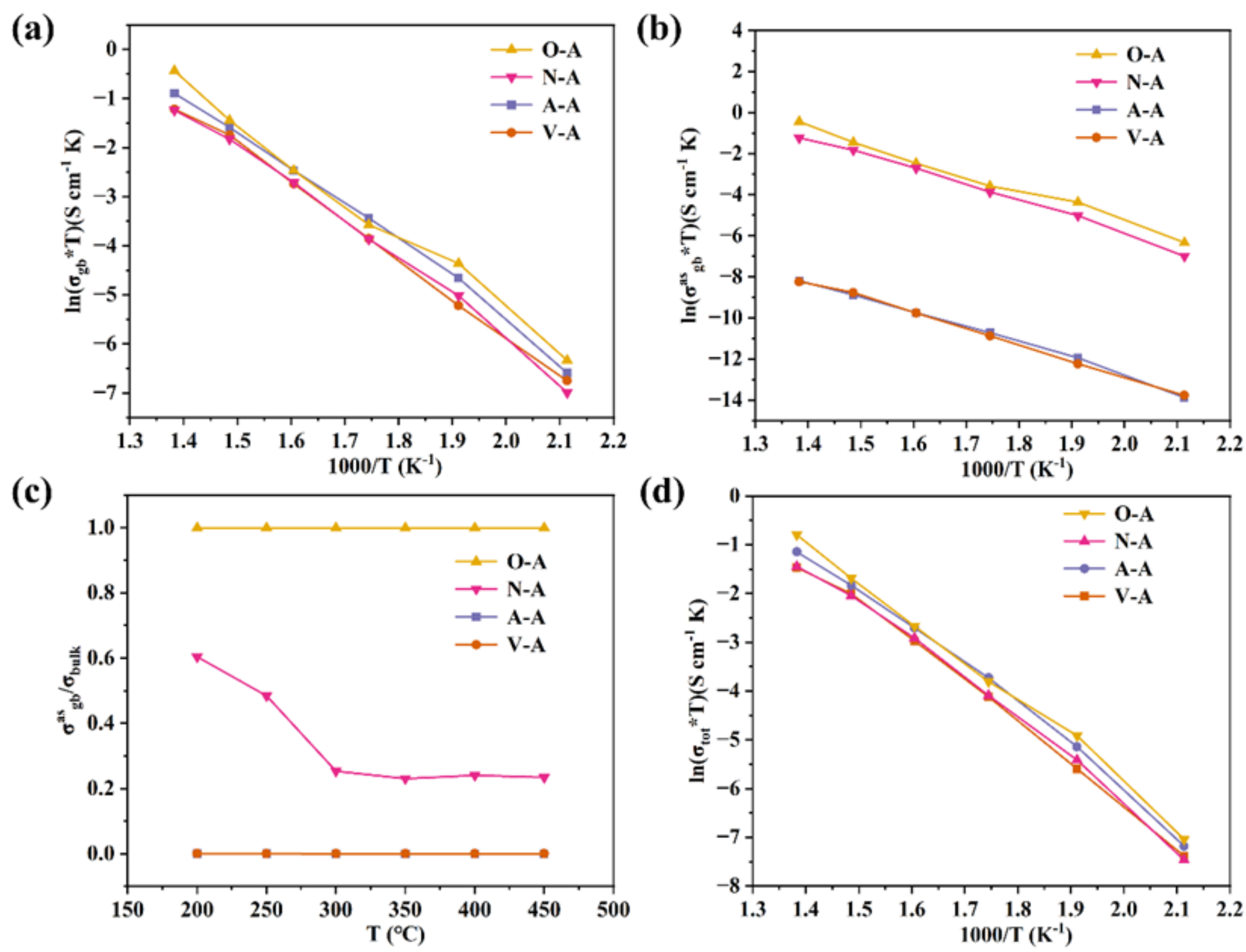

**Figure 6.** Arrhenius plots of (a) grain-boundary conductivity, (b) apparent grain-boundary conductivity, (c) conductivity after correction for the bulk contribution, and (d) total conductivity of $Na_{0.5}Bi_{0.465}Sr_{0.02}Eu_{0.005}TiO_{3-\delta}$ ceramics pre-calcined under different oxygen partial pressures.

The brick-layer model was used to analyze the intrinsic grain-boundary conductivity. As shown in the apparent grain-boundary conductivity plot in Figure 6b, the grain-boundary conductivities can be divided into two groups: the high-conductivity N-A and O-A samples and the low-conductivity A-A and V-A samples. The O-A sample has the smallest average grain size and therefore the highest grain-boundary density; after excluding the influence of grain-boundary number, it still shows superior conductive behavior. The effect of bulk conductivity was subsequently corrected, as shown in Figure 6c, allowing the independent contribution of grain-boundary structure to the conduction performance to be separated and quantified. The influence of impurity coverage on grain-boundary conductivity was also considered. In the brick-layer model, cubic grains are surrounded by grain-boundary layers of equal thickness, and $(\omega/dg)^2$ represents the impurity-blocking term, which depends on grain size and impurity distribution. According to the SEM results, the N-A sample has the largest grains. The structural analysis indicates relatively high phase purity, so the impurity content of the different samples can be regarded as comparable and very low. When grain growth is nearly complete, grain-boundary impurity coverage decreases owing to the scavenging and dilution effects at grain boundaries. For the total conductivity, the O-A sample shows the highest value.

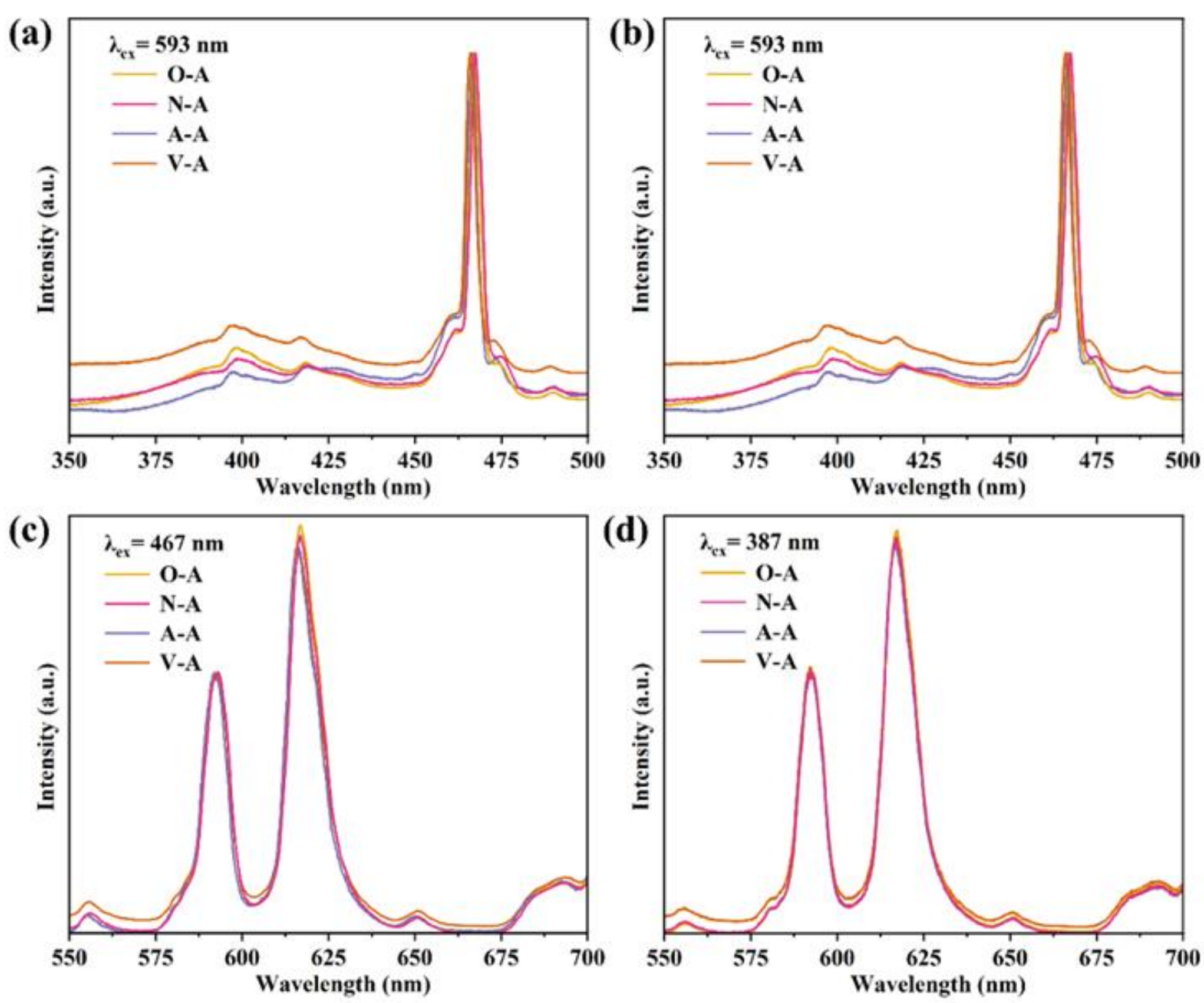

**Figure 7.** PLE spectra of $Na_{0.5}Bi_{0.465}Sr_{0.02}Eu_{0.005}TiO_{3-\delta}$ ceramics monitored at (a) the magnetic-dipole and (b) electric-dipole emissions, and PL spectra under (c) intrinsic excitation and (d) charge-transfer (CT) excitation.

To elucidate the effect of heat-treatment atmosphere on the local structure of $Na_{0.5}Bi_{0.465}Sr_{0.02}Eu_{0.005}TiO_{3-\delta}$ ceramics, excitation spectra monitored at 593 and 616 nm and emission spectra under different excitation wavelengths were recorded, as shown in Figure 7. Figures 7a and 7b show the excitation spectra monitored at 593 and 616 nm, respectively, normalized to the 467 nm band. No obvious shift of the CT band is observed. Analysis of the I_CT/I_4f ratio shows that the O-A sample has the largest value. This may arise because oxidation modifies the electron-cloud density and polarizability of lattice oxygen ions. Even with fewer oxygen vacancies, electrons around oxygen are more readily excited and transferred to the rare-earth center, thereby enhancing the charge-transfer process. The relatively large ratio in the V-A sample has a different origin: the high concentration of oxygen vacancies produced under low oxygen partial pressure distorts the lattice and may even lead to $V_{na}'$-$V_{o}^{\bullet\bullet}$-$V_{na}'$ defect associates, resulting in severe local distortion. An appropriate balance in metal-oxygen bonding favors charge transfer while maintaining structural stability [16–18].

Figure 7c shows the emission spectra under 467 nm excitation, normalized to the magnetic-dipole emission. Analysis of the electric-dipole intensity indicates that the O-A sample has the largest asymmetry ratio. Considering the efficient charge transfer observed in the excitation spectra, the transferred energy may preferentially populate $Eu^{3+}$ sites with lower local symmetry, producing a larger macroscopic asymmetry ratio. In the V-A sample, a high concentration of oxygen vacancies causes lattice distortion, but it may also partially reduce $Eu^{3+}$ to $Eu^{2+}$ and create disordered $Eu^{3+}$ sites, leading to broadened emission bands. The lower asymmetry ratio observed for this sample is therefore attributed to the competition between these two effects. CT excitation reflects the overall structural environment of the sample. The O-A sample again shows a clearly larger asymmetry ratio, whereas the other spectra nearly overlap. This confirms that CT excitation does not selectively probe the most distorted regions but instead provides a statistically averaged description of the $Eu^{3+}$ local environment. Severe disruption of long-range lattice periodicity increases the diversity of local environments and corresponds to highly asymmetric coordination [14], which ultimately leads to conductivity degradation.

## Conclusions

The influence of heat-treatment atmosphere on the structure and electrical properties of $Na_{0.5}Bi_{0.465}Sr_{0.02}Eu_{0.005}TiO_{3-\delta}$ ceramics was investigated. Oxygen partial pressure regulates grain size and structural ordering through three coupled mechanisms: Bi volatilization, grain-boundary oxygen adsorption, and oxygen-vacancy association. Under high oxygen partial pressure, although the overall oxygen-vacancy concentration is the lowest, isolated oxygen vacancies, a low migration activation energy (0.68 eV), and minimal grain-boundary impurity coverage together lead to the highest bulk, grain-boundary, and total conductivities. Low oxygen partial pressure promotes severe Bi volatilization, A-site defect association, and partial $Eu^{3+}$ reduction, increasing the activation energy to 0.88 eV and decreasing the conductivity by approximately one order of magnitude. The $Eu^{3+}$ asymmetry ratio shows a strong positive correlation with bulk conductivity and can serve as an optical probe for evaluating the effective mobility of oxygen vacancies. The order-disorder competition and grain-boundary migration-pinning competition proposed here provide a general framework for optimizing perovskite-type oxide-ion conductors by atmosphere engineering.

## Author Contribution Statement

Zongxue Wang: performed the experiments, collected and analyzed the data, and drafted and revised the manuscript.

Duanting Yan: conceived the research idea and designed the research plan.

Hancheng Zhu: revised and approved the final version of the manuscript.